\documentclass[aps,prb,showpacs,showkeys,twocolumn,amsmath,amssymb,reprint]{revtex4-1}

\usepackage{times}
\usepackage{color}
\usepackage{graphicx}
\usepackage{amsmath,amsbsy,amsfonts,mathrsfs}

\newcommand{\be}{\begin{equation}}
\newcommand{\ee}{\end{equation}}
\newcommand{\bea}{\begin{eqnarray}}
\newcommand{\eea}{\end{eqnarray}}
\newcommand{\bg}{\begin{figure}}
\newcommand{\eg}{\end{figure}}
\newcommand{\bi}{\begin{itemize}}
\newcommand{\ei}{\end{itemize}}

\usepackage[version=3]{mhchem} 
\usepackage{dcolumn}
\usepackage{textcomp}

\begin{document}
\bibliographystyle{apsrev}


\title[Hydrocarbons and Hydrocarbyls in Space]{
Formation of Hydrocarbons from Hydrogenated Graphene in Circumstellar Clouds}

\author{\bf Jos\'e I. Mart\1nez}
\email{joseignacio.martinez@icmm.csic.es}
\address{Dept. Surfaces, Coatings \& Molecular Astrophysics, Institute of Materials Science of Madrid (ICMM-CSIC), ES-28049 Madrid (Spain)}

\author{\bf Jos\'e A. Mart\1n-Gago}
\address{Dept. Surfaces, Coatings \& Molecular Astrophysics, Institute of Materials Science of Madrid (ICMM-CSIC), ES-28049 Madrid (Spain)}

\author{\bf Jos\'e Cernicharo}
\address{Dept. Surfaces, Coatings \& Molecular Astrophysics, Institute of Materials Science of Madrid (ICMM-CSIC), ES-28049 Madrid (Spain)}

\author{\bf Pedro L. de Andres}
\address{Dept. Surfaces, Coatings \& Molecular Astrophysics, Institute of Materials Science of Madrid (ICMM-CSIC), ES-28049 Madrid (Spain)}

\date{\today}

\begin{abstract}
We describe a mechanism that explains the formation of hydrocarbons and hydrocarbyls from hydrogenated graphene/graphite; hard C--C bonds are weakened and broken by the synergistic effect of chemisorbed hydrogen and high temperature vibrations. Total energies, optimized structures, and transition states are obtained from Density Functional Theory simulations. These values have been used to determine the Boltzman probability for a thermal fluctuation to overcome the kinetic barriers, yielding the time scale for an event to occur. This mechanism can be used to rationalize the possible routes for the creation of small hydrocarbons and hydrocarbyls from etched graphene/graphite in stellar regions.
\end{abstract}




\maketitle


Recent technological advances providing improved sensitivity in radio-astronomical receivers, and through the usage of large (sub)millimeter radio telescopes, have permitted to find numerous new chemical species in outer space. Several chemical processes lead to the formation of these molecules in gas-phase in space; 
e.g. ion-neutral, neutral-neutral, and radical-neutral reactions. In addition, reactions at the surface of the grains can also enhance the gas-phase abundance of these species when the temperature of dust grains increases above a critical value. 
Heating processes of dust grains can occur due to shocks, UV photon absorption from external sources, or by the proximity of a newly formed star. 
Nevertheless, the presence of complex molecules in photodissociation regions requires a continuous production mechanism that so far is not fully understood,
although undestanding these chemical processes is crucial to study the chemical evolution of the interstellar gas in galaxies. 
Therefore, there is an increasing interest in detailed microscopic calculations leading to models that can explain the relative abundance of these molecules. 

Carbonaceous particles are usually terminated on graphene/graphite-like layers or amorphous carbon, which may transform under heating to other stable forms of carbon. 
Clouds rich on silicates or carbonaceous dust grains 
are covered with ices (\ce{H2O}, \ce{NH3}, \ce{CH4}, \ce{CH3OH}, \ce{CO}, $\ldots$) 
providing support for the formation of different species. A paradigmatic example
is the formation of the hydrogen molecule from atomic hydrogen;
due to the presence of ices it is produced even at temperatures as low as 10 K, as 
demonstrated by laboratory experiments on circumstellar grain analogs and
by Density Functional Theory (DFT) simulations.\cite{hornekaer03} 
On the other hand, in low-density clouds (diffuse) these particles are known to be {\it clean} of contaminants like ice, and a similar mechanism is lacking since the thermally activated process of molecular hydrogen formation would require temperatures $T>250$ K to achieve a reasonable yield in times ranging between a few years to some thousands years.\cite{mckay,CasoloJPCA2009}


\begin{figure}
\centerline{\includegraphics[width=0.99\columnwidth]{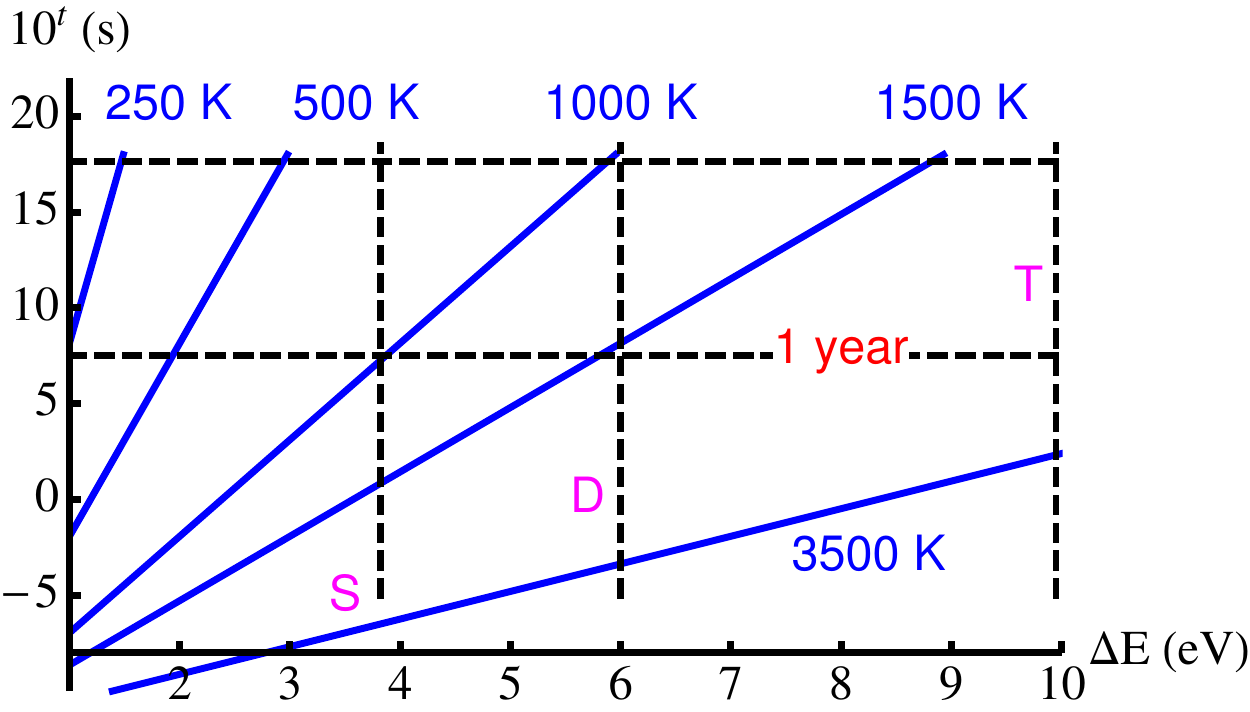}}
\smallskip \caption{(color online). Continuos (blue) lines give typical times ($10^t$ in s) to overcome an energy barrier 
($\Delta E$ in eV) by picking up a thermal fluctuation at a temperature $T$ (in K) between $250$ and $3500$ K. Characteristic energies 
for the simple (S), double (D) and triple (T) C--C bonds are represented by vertical dashed lines. The top dashed horizontal line 
represent the total time since the big-bang ($\approx 10^{10}$ years), and the middle horizontal line represents 1 year.
Around the blue $3500$ K line the breaking of C--C bonds in the solid-liquid phase transition
takes place in about ms.
\label{fgr:boltzmann}} 
\end{figure}

The opposite problem is the effect of atomic hydrogen on carbonaceous particles.\cite{GuisingerNL2009} 
It has recently 
attracted attention because 
of its potential to explain the formation of hydrocarbon molecules and 
all sort of radicals in space.\cite{merino} 
These authors have described the 
high-temperature etching of \ce{SiC} grains in a hydrogen-rich atmosphere in the laboratory. 
The main experimental observation is the
formation of holes on a graphene surface. 
This is surprising since the typical energy stored in a 
C--C bond on graphene is $\ge 4$ eV, and such a 
process would only happen on a time-scale of several years in pristine graphene at 
the temperature of the experiment ($1000$ K), cf. Fig.~\ref{fgr:boltzmann}. 
Indeed, graphene is one of the known toughest materials in Nature, with a very large in-plane Young modulus of $\approx 1$ TPa, which arises from the extraordinary 
efficiency of the sp$^2$ and $\pi$-like hybrid orbitals bonding C atoms on a honeycomb topology.\cite{deandres12b} 
Furthermore, bonding between C atoms is very versatile, 
resulting in a range of strengths from triple to single bonds, 
$\approx 10$ eV to $4$ eV per bond, respectively.
Characteristic energies for the simple (S), double (D) and triple (T) carbon-carbon bond are represented by vertical dashed lines in Fig.~\ref{fgr:boltzmann}.\cite{langhoff}
Graphene is located on an intermediate region with a coordination order of about $1\frac{1}{3}$ that corresponds to $\approx 4-5$ eV per C--C bond. 
Such a kinetic barrier for the breaking of C--C bonds is too high for the 
process to take place at $1000$ K on the observed experimental time scale, i.e. minutes. 
Owing to the exponential dependence with temperature of the Boltzmann factors
such a process at $500$ K would take an amount of time
equivalent to the Age of the Universe and would not happen,
while at the same time the solid-liquid phase-transition starting around
$3500$ K is correctly predicted to take place in about ms. 
We notice that a reduction of the barrier below $3$ eV would be enough to allow such a 
process to happen on about a few minutes.
Experiments run on heavily hydrogenated graphene make it clear that the presence of hydrogen 
does not eliminate these kinetic
barriers\cite{Elias009}, but we find that it acts to reduce their values so it can happen on a reasonable laboratory time scale, between miliseconds and minutes.\cite{merino} 
Here we search for detailed mechanisms that can explain the {\it catalytic}-like activity of atomic hydrogen leading to the 
reported formation of holes in the graphene layer, as well as the nature of the fragments formed at the same time, i.e. hydrocarbons and radicals of various sizes. 


In the {\it ab-initio} atomistic simulations, total energies, forces and stresses were minimized by using  DFT as implemented in the plane-waves package 
{\sc Quantum Espresso}.\cite{PWSCF} 
A perturbative van der Waals (vdW) correction was used to check for the effect of long-range interactions on the different configurations analyzed. For this purpose, we have used an empirical vdW R$^{-6}$ correction to add dispersive forces to conventional density functionals within the DFT+D formalism.\cite{Grimme,Elstner,Gavezzotti} This vdW contribution to the total energy is almost negligible around the transition states; it only becomes significative ($\sim$15\%) when distances between the expelled fragments and graphene increases $>$ 4\AA, which only matters for a good description of energies at asymptotic distances.
The Local Density Approximation (LDA) has been used for the exchange and correlation potential. 
It describes fairly well both the C--C and C--H bonds, and at the same time offers a most simple and clear conceptual frame. Similar conclusions have been reached by checking a few selected cases
with a 
Generalized Gradient Approximations (GGA) to the exchange and correlation potential. 
Norm-conserving scalar-relativistic pseudopotentials have been used to model the ion-electron 
interaction.\cite{Vanderbilt}

Models having from tens to hundreds of atoms have been considered to construct both finite-size clusters ({labeled as \bf C}) and extended periodic-cell models (labeled as {\bf c}). A {\bf k}-space mesh of $\Delta {\bf k} \le 0.01$ {\AA}$^{-1}$, and an energy cutoff of $500$ eV yielded total energies with an accuracy of $\Delta E \approx \pm 0.05$ eV (converged to a precision better than $10^{-6}$ eV). Optimized geometries were obtained with residual forces lower than $0.01$ eV/{\AA}, and stresses below $0.1$ GPa. Besides enthalpies of formation, $H$, for a particular process the all important feature to ascertain its feasibility is the height of the barrier at the transition state (TS), $\Delta E$. 
These TSs have been investigated within the climbing-image nudge elastic band (CI-NEB) approach~\cite{NEB0,NEB1,NEB2} implemented in the {\sc Quantum Espresso} package,~\cite{PWSCF} where the initial, the final and all the intermediate image-states were free to fully relax. At this point, it is important to remark that the underlying {\it chemistry} of these systems is well represented by all the approximations adopted here, and we expect a good correlation with experimental data, as it is usually attained \cite{besenbacher09JACS, GrimmePCCP2006, ChenJPCC2007}.


\begin{figure}
\includegraphics[width=0.99\columnwidth]{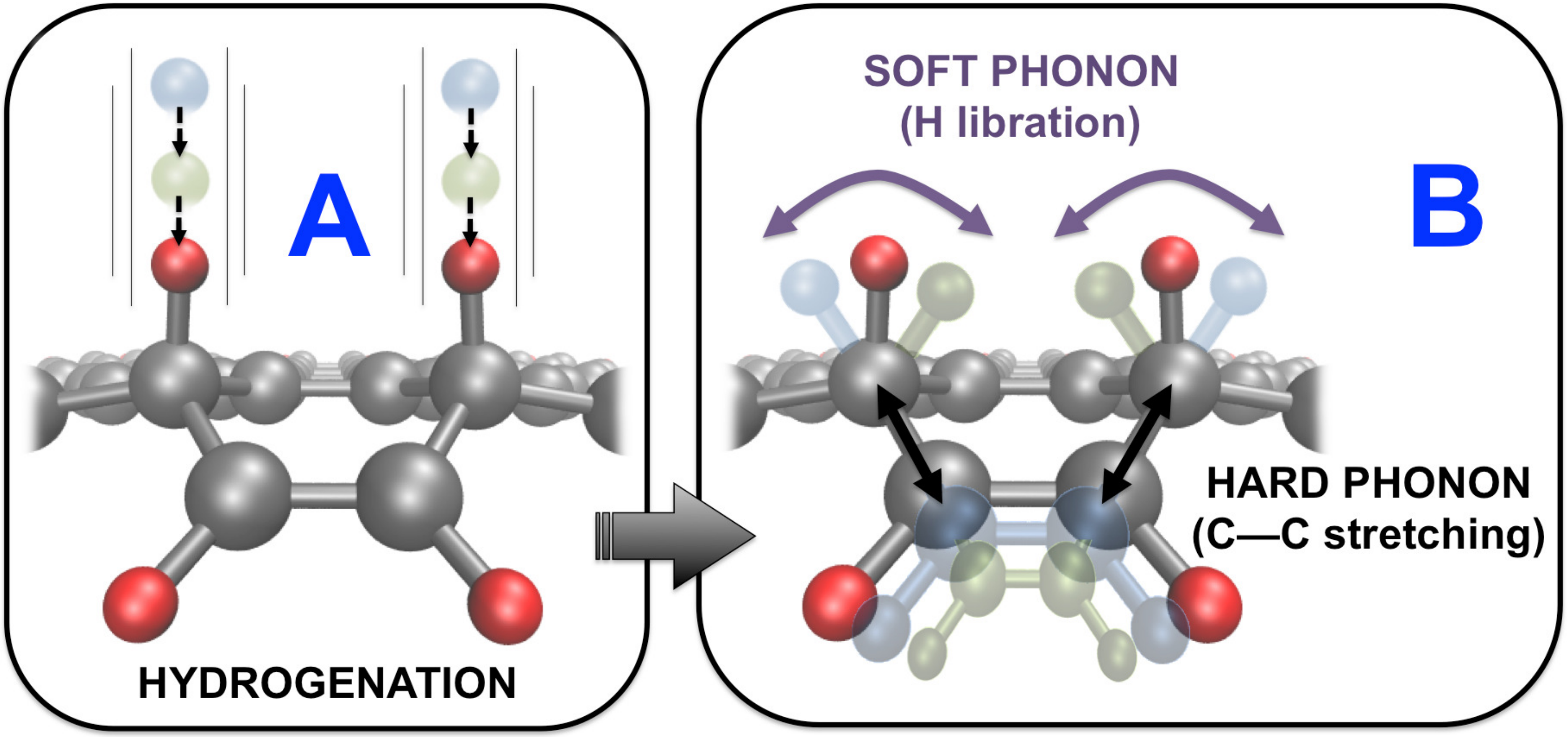}
\smallskip \caption{
(color online) Schematic mechanism to detach a \ce{C2H2} molecule from the edge of a hydrogenated graphene nano-ribbon.
In step A hydrogen atoms are chemisorbed on nearby positions to progress towards a transition state under the combination
of two phonons (step B). Once the H atoms have been inserted near the transition state the C--C bonds are weakened enough to
be broken by thermal activation after annealing.
\label{fgr:TheExp}} 
\end{figure}

A plausible mechanism that would allow the breaking of the strong in-plane C--C bonds should involve a transition state 
low enough to get from the initial to the final state in a meaningful time on the experiment time scale.
Our simulations show that atomic hydrogen plays a dual role in this job.
On the one hand, it chemisorbs on the graphene layer with an adsorption energy of $\approx 1$ eV, and a small initial sticking
barrier of $\approx 0.15$ eV. This small barrier appears as a result of the induced deformation upon adsorption of H on the stiff planar 
graphene layer, where the sp$^{2}$ acquires a partial sp$^{3}$ character \cite{BalgarJPCL2013, JohnsACR2013}. 
On the other hand, hydrogen can adsorb on the edges of the newly
created holes saturating the dangling bonds very effectively.
That kind of C--H bond can store as much as $5$ eV and it is characterized by vibrational frequencies that match the
phonons of the graphene layer.\cite{deandres12b} Therefore, the C--H bond can effectively compete with the C--C 
bond making it possible to reduce transition barriers for various processes to reasonable amounts (see Fig.~\ref{fgr:TheExp}). 
Fig.~\ref{fgr:boltzmann} gives in a logarithmic scale the typical time in seconds of having a process with a transition barrier of $\Delta E$ (eV) 
at a temperature $T$ (K). At the temperatures of interest ($T>500$ K) we can safely neglect tunneling contributions and we assume that the probability for a thermal fluctuation to overcome the TS barrier is  determined by the canonical Boltzmann factor,\cite{deandres93}
\be \label{eq1} 
\Gamma = \Gamma_0 e^{-\frac{\Delta E}{k_B T}}.
\ee Since the exponential dominates the behaviour of this probability as a function of $T$ we take the prefactor as a temperature-independent
value, which can be estimated via a typical C--C or C--H phonon, $\Gamma_0 \approx 1000$ cm$^{-1}$. 
Due to the Boltzmann factor, variations on $\Gamma_0$ by even a few orders of magnitude
are quickly overcome by the effect of the temperature. 
By propagation of errors, we find for typical temperatures of $\approx 1000$ K,
and typical energy barriers of $\approx 1$ eV that uncertainties in the calculation of barriers
of $\approx \pm 0.05$ eV, and in the values of temperatures of $\approx \pm 10$ K, 
result in Fig.~\ref{fgr:boltzmann} in fractional errors $< 35$\%. 
These errors bars, even much bigger values, are quite acceptable to draw conclusions since the rates change so
quickly with the values of the barriers.

\begin{figure}
\includegraphics[width=0.99\columnwidth]{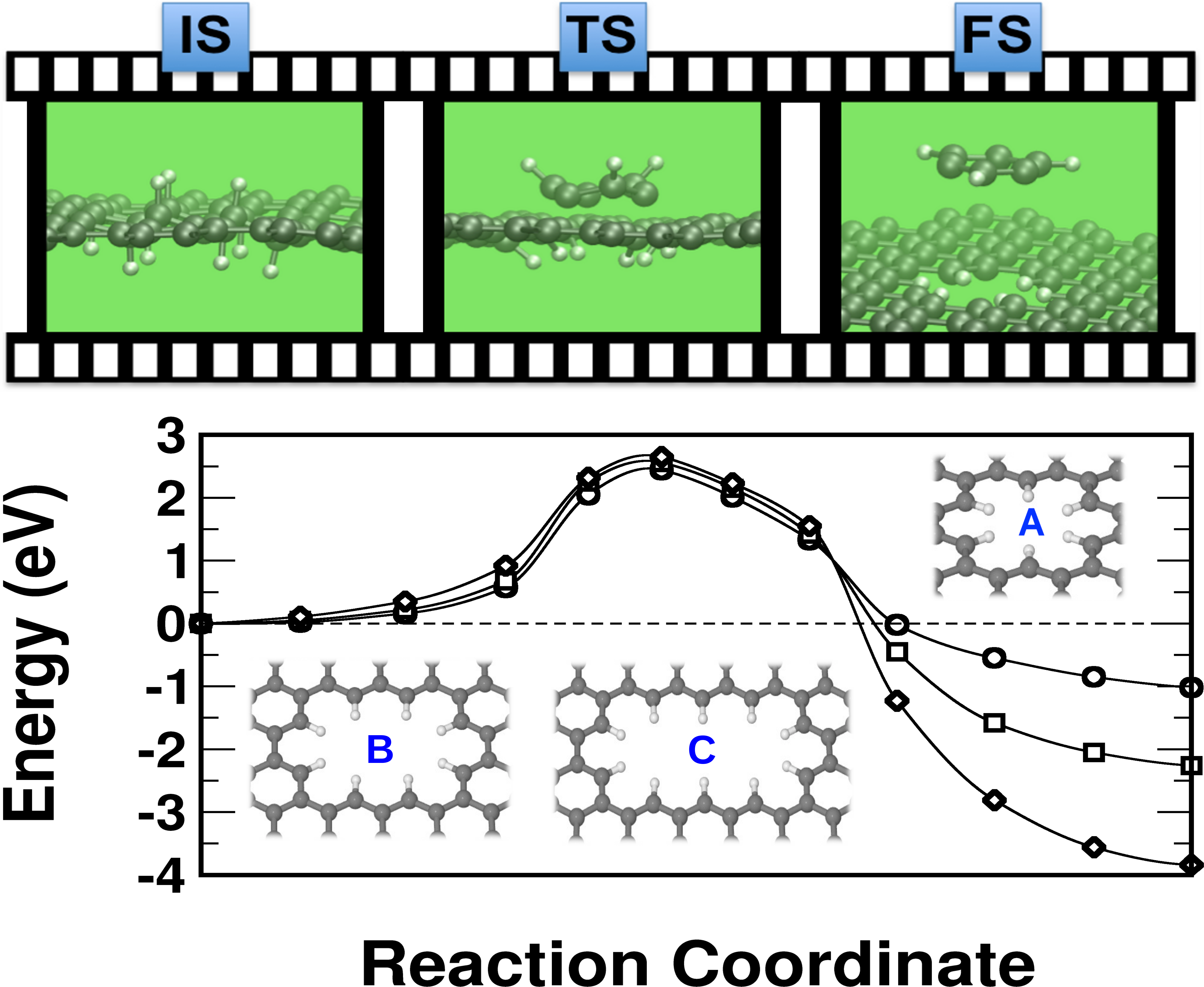} \\
\smallskip \caption{(color online) {\bf Top:} pictorial sketch of the initial (IS), transition (TS) and final (FS) states for the \ce{C6H3} 
dettachment. {\bf Bottom:} Energy barriers to create holes of different sizes on an extended periodic graphene layer. ($\circ$) 
corresponds to the creation of a \ce{C6} hole, ($\square$) to a \ce{C10} hole, and ($\diamond$) to a \ce{C14} hole. 
\label{fgr:TheHoleTh}} 
\end{figure}


\begin{table}
\caption[tbl:TheHoleI]{Perimetral length, $P$ (\AA), enthalpies of formation, $H$ (eV), and transition states barriers, $\Delta E$ (eV), for different sizes of the hole
(created by dettaching \ce{C6H3}, \ce{C10H4} and \ce{C14H5}). {\bf C} and {\bf c} indicate finite-size clusters and periodic-cell models, respectively. \label{tbl:TheHoleI}}
\begin{tabular*}{\columnwidth}{@{\extracolsep{\fill}}lc|ccc|cccccc|cccccc}
\hline \hline
            	&&&$P$(\AA)&&&& $H$ (eV) &&&&&& $\Delta E$ (eV) & \\ \hline \hline
\text{\bf Species}  & & & & & & {\bf C} & & & {\bf c} & & & {\bf C} & & & {\bf c}           
\\ \hline
\ce{C6H3}            	&&& 9	&&&  -0.82	&&& -1.12  &&& 2.81 	&&& 2.43  &   	\\
\ce{C10H4}         	&&& 15 	&&&  ---	&&& -2.35  &&& --- 	&&& 2.56  &   	\\
\ce{C14H5}         	&&& 21	&&&  ---	&&& -3.96  &&& ---  	&&& 2.65  &   	
\\ \hline \hline
\end{tabular*}
\end{table}


First, we discuss how a {\it defective} hole on a graphene layer can be created, 
as it has been observed in UHV-STM and AFM experimental images.\cite{merino} 
We start by adsorbing a few H atoms on both sides of the graphene layer
surrounding the region where the hole may appear and producing a sp$^{3}$-like
deformation on the layer. 
The role of chemisorbed atoms is to weaken first, and then to passivate the broken C--C bonds.  
Take as an example the extraction of a six C ring (see Fig. \ref{fgr:TheHoleTh}). 
It would be desirable from the point of view of the formation enthalpy
to saturate this ring with H atoms to directly obtain benzene. However, this process
involves too high barriers due to steric impediments from hydrogens in the extracted fragment
and other hydrogens left in the layer. Instead, we find it is optimum to extract \ce{C6H3}.
The process starts by breaking the three C--C bonds that are left unsaturated in the fragment,
and subsequently the three remaining ones. 
In the absence of chemisorbed hydrogen the formation of such a six-C ring hole has a 
barrier of $+7.12$ eV, which at $1000$ K simply would not happen even in typical stellar times,
Fig.~\ref{fgr:boltzmann}.
Adsorption of atomic hydrogen drastically modifies this picture; 
values in Table~\ref{tbl:TheHoleI} show how the transition state 
to extract \ce{C6H3} is reduced 
down to $+2.43$ eV, a process that now may take place in less than a second at $1000$ K. 
The mechanism involves C--C stretching so a wagging mode of a nearby 
chemisorbed H can place the H closer to the middle of the C--C bond and can help to break it while 
forming at the same time a new C--H bond, cf. Fig.~\ref{fgr:TheExp}. 
The formation of a favourable configuration of chemisorbed hydrogen atoms depends on the available
density of atoms impinging on the surface; 
the reference experiments have been performed under conditions
where atomic H has been admitted in the UHV chamber
up to a $10^{-6}$ mbars ($\approx 10^{10}$ atoms/cm$^3$), and
temperature has been raised up to $1000$ K.\cite{merino}
These conditions imply hydrogen saturation conditions, making likely
to reach a favourable configuration, that then determines the shape of the hole.
These experiments have been designed to mimick regions near red-giant stars 
between $3$ to $15$ stellar radii, where temperature ranges from 
$1500$ K to $200$ K, and hydrogen densities take values between
$10^{11}$ to $10^{6}$ atoms/cm$^{3}$.\cite{agundez}
We remark that the enthalpy of formation for this particular
\ce{C6} hole is negative ($-1.12$ eV), therefore making
the hole saturated with hydrogen and the fragment, \ce{C6H3}, stable. 
A similar pattern follows for a large cluster labelled {\bf C} in 
Table~\ref{tbl:TheHoleI} with values of $+2.81$ and $-0.82$ eV for the barrier and the enthalpy of formation, respectively. 
The main difference between the periodic and finite-size simulations resides in the barriers to stick the hydrogen initially near the region to be etched. This is mainly due to the different elastic constants for a continuous layer and a small chunk of material that may be supported by some substrate or not. 
Finally, we analyze, for the case of the periodic system, the upstart of the scaling with the size of the hole formed, i.e. with the number of C atoms to be extracted, by considering 10 and 14 C atoms. 
These holes have been labeled in Fig.~\ref{fgr:TheHoleTh} with A, B and C, and in the continous lines marked with empty circles, squares and diamonds, respectively. 
Enthalpies of formation grow slowly with the size of the fragment tending to the saturation value to be found on very large systems. 
Interestingly enough, however, the transition state barriers tend to be nearly the same, indicating that the main difficulty to extract large fragments resides with the configurational statistical complexity more than with the chemistry. This can be fixed by introducing an abundant amount of hydrogen favoring 
the appearance of many different configurations; the holes observed would then correspond to the particular configuration realized on the graphene layer.
Finally, we notice that in an elegant experiment Xie et al. have recently measured the etching of
single graphene layer edges by hydrogen plasma at a rate of approximately
$0.3$ nm/min.\cite{XieJACS2010}
According to our model, to create a \ce{C6} hole with a perimetral length of approximately $9$ {\AA} at the
measured rate it would require a barrier of $\approx 2.85$ eV, which is remarkably close to
the values in Table I taking into account that our focus is on the general 
mechanism using arguments based in the order of magnitude of rates, not on the detailed 
values that we cannot expect to obtain with such accuracy. 

\begin{figure}
\includegraphics[width=0.99\columnwidth]{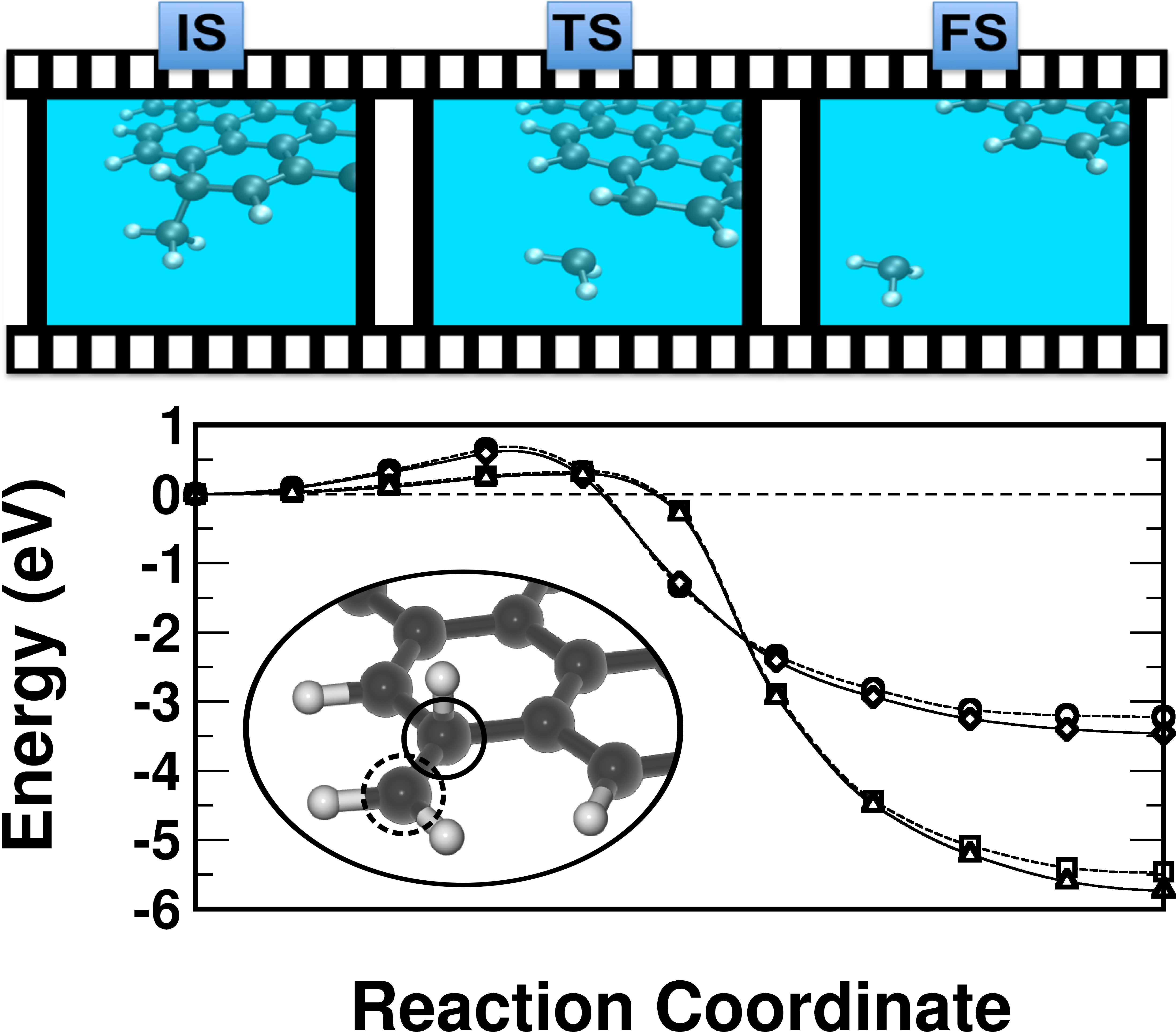} \\
\smallskip \caption{(color online) {\bf Top:} pictorial sketch of the initial (IS), transition (TS) and final (FS) states for the \ce{CH3} 
dettachment. {\bf Bottom:} Energy barriers for the proposed mechanism of breaking a C--C bond in the edge of a H saturated graphene nano-ribbon. A chemisorbed H on top a C atom (continuous circle, referred in the text as the first C) moves towards a transition state located in between the second C (dashed circle) to create fragments like \ce{CH2} (inset) or \ce{CH3} (upper pannel). ($\circ$,$\square$) \ce{CH2} and \ce{CH3} from a H saturated edge; and
($\diamond$,$\bigtriangleup$) \ce{CH2} and \ce{CH3} from a H saturated hole. \label{fgr:TheFragmentsTh}} 
\end{figure}

\begin{table}
\caption[tbl:PAH]{Enthalpies of formation, $H$ (eV), and transition states barriers, $\Delta E$ (eV), to obtain different fragments from the inner
part of the hole in the graphene layer. {\bf C} and {\bf c} indicate finite-size clusters and periodic-cell models, respectively. \label{tbl:PAH}}
\begin{tabular*}{\columnwidth}{@{\extracolsep{\fill}}lc|cccccc|cccccc}
\hline \hline
            	&&&& $H$ (eV) &&&&&& $\Delta E$ (eV) & \\ \hline \hline
\text{\bf Molecule}  & & & {\bf C} & & & {\bf c} & & & {\bf C} & & & {\bf c}           
\\ \hline
CH$_2$           	&&&  -3.22	&&& -3.46  	&&&   0.65 	&&& 0.59  &   	\\
CH$_3$         	&&&  -5.46	&&& -5.73  	&&&   0.33 	&&& 0.29  &   	
\\ \hline
C$_2$H$_2$   	&&&  +1.52	&&& +1.33  	&&&   2.35  	&&& 2.18  &   	\\
C$_2$H$_4$   	&&&  +0.81	&&& +0.74	&&&   1.93 	&&& 1.79  &   	\\
C$_2$H$_6$   	&&&  -0.61	&&& -0.82  	&&&   1.48  	&&& 1.67  &   	
\\ \hline \hline
\end{tabular*}
\end{table}


Second, we focus on the small molecules formed on the internal edges of the holes left on the
surface, and in the external edges of the ejected fragments following the
creation of a large defect on the layer. 
Table~\ref{tbl:PAH}
shows a comparison between enthalpies and transition state barriers for a variety of small molecules. 
Formation of \ce{CH2} and \ce{CH3} involves the lowest barriers and the most stable systems, 
e.g. see discontinuous lines in Fig.~\ref{fgr:TheHoleTh}. These two molecules 
can be formed on the external edges of graphene flakes, or on the internal edges of the hole ({\bf c}). As an example of an straightforward mechanism on the edge of a hydrogen saturated nano-ribbon we draw an external C atom (surrounded by a dashed circle) attached to two 
H atoms and a third H chemisorbed on a top position in the nearby C (surrounded by a continuous circle). Barriers to form both \ce{CH2} and \ce{CH3} are similar since the important point here is the local geometry giving raise to the breaking of a C--C bond by insertion of a chemisorbed H. The more saturated with hydrogen the fragment, the lower are the barriers, and the formation enthalpies are more negative, since the C--C bond becomes weaker. This mechanism does not result in the formation of \ce{CH} nor \ce{CH4}, although for opposite reasons. In the first case, the H chemisorbed on top position (continuous circle) finds it too favorable to bond to the second C (dashed circle) and cannot participate in the breaking of the C--C bond. In the second case, the second carbon atom cannot support the interaction with four H plus the first carbon atom. Nevertheless, \ce{CH2} and \ce{CH3} fragments dettached from hydrogenated graphene/graphite will capture environmental H atoms to spontaneously form \ce{CH4}.\cite{WoodJPC69}
Other fragments with higher C content, such as \ce{C2H2}, \ce{C2H4}, and \ce{C2H6}, are not exothermic, but display reasonable formation barriers too. 
The inverse process, where the fragment joins the edge, 
is hindered by entropic orientational considerations and does not play an important role. 
The transition barriers for these molecules having two C atoms are higher than for those having just one C atom, but still reasonable if compared with typical times in the circumstellar medium. To complement the scenario, since these particles may be bombarded by charged particles in a realistic situation located near a star we have considered the breaking of \ce{C2H2} on a charged surface, leading to the anion \ce{C2H2-}. The barrier for this case is of $+2.20$ eV, i.e. very similar to the one found on the inner side of the hole for the periodic system. The presence of the extra charge on the surface does not significantly alter the mechanism of breaking the relevant C-C bonds, and indicates that the abundance of anions and neutral species should be similar. Since the experiments have been performed on graphene formed on a \ce{SiC} substrate we have additionally tested the influence of the layer immediately below the surface, the so-called buffer layer. For \ce{C2H2} we see a negligible change of the height of the barrier, from $2.35$ eV to $2.33$ eV. This finding manifests that the weak interaction between layers do not interfere with the strong chemical-like bonds broken and formed that explains the mechanism for etching of the layer, and assigns to the buffer layer a purely physical confining role \cite{BalgarJPCL2013,SeyllerPRB2005}. Since in most cases graphene layers grown on metallic surfaces keeps also a weak interaction with the support, we expect the same principles to apply for those cases.



Summarizing, in this letter we propose a mechanism underlying the formation of prototypical hydrocarbons, hydrocarbyls, and other polyaromatics from hydrogenated graphene. 
We have studied the formation of large holes on graphene upon atomic hydrogen adsorption and high-T annealing, as observed in experiments performed in the laboratory.
These have been rationalized in terms of the adsorption of H atoms around the region where the hole is expected to appear. Chemisorbed atoms act to passivate broken bonds in synergy with phonons simultaneously weakening the hard C--C bonds. 
This is a temperature-activated process exhibiting typical barriers of around $3$ eV, and occurs at $1000$ K in a few minutes. Computed transition state energies are used to determine the Boltzman probability for a thermal fluctuation to overcome such a barrier, 
and the time scale for an event to occur, which is of paramount importance from the practical point of view. The proposed mechanism also sheds light into possible routes for the creation in the circumstellar medium of small hydrocarbons and hydrocarbyls, such as \ce{CH2}, \ce{CH3}, \ce{C2H2}, \ce{C2H2-}, \ce{C2H4} and \ce{C2H6}.

We acknowledge funding from the Spanish MINECO (Grants MAT2011-26534 and CSD2009-00038), the EU (FP7 Program Grant 604391, and ERC Synergy Grant ERC-2013-SYG-610256), and computing resources from CTI-CSIC. JIM acknowledges a CSIC-JaeDoc Fellowship, cofunded by ESF.

\bibliography{bib} 

\end{document}